\documentstyle[12pt]{article}

\def\be{\begin{equation}}
\def\textlineskip{}
\def\bibit{\it}

\def\bibbf{\bf}
\def\ee{\end{equation}}

\def\ba{\begin{array}{c}}
\def\ea{\end{array}}

\begin{document}

\titlepage
\vspace*{2cm}

\begin{center}{\Large \bf
Spontaneous breakdown of ${\cal PT}$ symmetry

in the solvable square-well model }\end{center}

\vspace{10mm}

\begin{center}
Miloslav Znojil \vspace{3mm}

Nuclear Physics Institute of Academy of Sciences of the Czech
Republic, 250 68 \v{R}e\v{z}, Czech Republic

e-mail: znojil@ujf.cas.cz

\vspace{3mm}

and

\vspace{3mm} G\'{e}za L\'{e}vai

\vspace{3mm}

Institute of Nuclear Research of the Hungarian Academy of
Sciences, PO Box 51, H-4001 Debrecen, Hungary

e-mail: levai@atomki.hu

\end{center}

\vspace{5mm}

\section*{Abstract}
Apparently, the energy levels merge and disappear in many ${\cal
PT}$ symmetric models. This interpretation is incorrect: In
square-well model we demonstrate how the doublets of states in
question continue to exist at complex conjugate energies in the
strongly non-Hermitian regime.

\vspace{5mm}

PACS {03.65.Fd,03.65.Ge,03.65.Nk}

\newpage

\vspace*{1pt}\textlineskip
\section{Introduction}
\vspace*{-0.5pt} \noindent
 In their pioneering work which dates
back to the late seventies, Caliceti et al\cite{Caliceti} have
noticed that, rather unexpectedly, non-Hermitian Hamiltonians of
certain type could generate the resonances with the {\em strictly
vanishing} width (i.e., the stable bound states with {\em strictly
real} energies). Rigorously (using a summability of perturbation
series), they proved that the energy levels remain real for the
sufficiently weak cubic anharmonicity {\em with the purely
imaginary coupling constant}. This result remained virtually
unnoticed even after Buslaev and Grecchi\cite{BG} discovered,
independently, that also the spectra of certain non-Hermitian
versions of the more current quartic anharmonic oscillators remain
real and bounded below even far beyond the mere perturbative
regime.

The decisive and abrupt increase of interest in all the similar
non-Hermitian Hamiltonians (characterized, usually, by the
property $H = {\cal PT}H{\cal PT}$ of the so called ${\cal PT}$
symmetry, with the parity ${\cal P}$ and the complex conjugation
${\cal T}$) has only been initiated by Bender and Boettcher in the
late nineties\cite{BB}. Using the so called delta expansions and
quasi-classical techniques they emphasized the possible relevance
of this particular type of the non-Hermiticity for the
phenomenological physics and field theory\cite{BM,BMa}. They also
conjectured\cite{BBM} the possible existence (and mathematical
consistence) of an alternative, ${\cal PT}$ symmetric quantum
mechanics.

The most important (and, sometimes, counterintuitive) properties
of the latter quickly developing formalism are most easily
clarified via solvable examples\cite{cjt98}-\cite{levai}. Within
this framework, special attention has been paid to the existence
and properties of the critical strengths of the interaction at
which the ${\cal PT}$ symmetry becomes spontaneously broken
\cite{Mandal,Ahmed}. One of the conclusions seems to be an
unexplained difference between the numerical and exactly solvable
models. In the former case (illustrated, say, on the popular
quartic anharmonic oscillator in ref.\cite{Simsek}), the breakdown
of the ${\cal PT}$ symmetry seems to take place at more steps.
This means that there exist many separate critical couplings
$Z^{(crit)}_n$ where at most a finite number of the energies
``merges" and disappears.

In the light of our recent study\cite{subm} the pattern looks
different for many exactly solvable models. We found that within
their shape-invariant subclass, the critical coupling
$Z^{(crit)}_n$ proves {\em always $n-$independent} and remains
unique. In the other words, all the energy levels stay real up to
the point beyond which we find no real energy at all.

Our present short note offers the resolution of the latter
apparent puzzle which proves to be an artifact of the choice of
the class of models in ref.\cite{subm}. We shall pick up a
different solvable model (viz., the ${\cal PT}$ symmetric ``square
well" of ref.\cite{sqw}) and show that the set of its critical
points $Z^{(crit)}_n$ remains very large (presumably, infinite).

An important additional merit of our new construction of solutions
lies in its extremely elementary nature. We would like to
emphasize that as an illustrative example, the ${\cal PT}$
symmetric square well proves at least as relevant as its standard
Hermitian predecessor. We would expect, in particular, that the
exceptional transparency of its solutions with broken symmetry
could clarify several open problems which still exist within
${\cal PT}$ symmetric quantum mechanics involving, e.g., the
necessary modification of the concept of Hermiticity
\cite{Krakov,Ali} and of the norm\cite{what,Quesne} as well as the
pseudo-unitarity of the time evolution\cite{pseudo} and the
various Sturm-Liouville oscillation theorems\cite{Sturm}.

\vspace*{1pt}\textlineskip
\section{Wave functions with broken ${\cal PT}$ symmetry}
\vspace*{-0.5pt} \noindent
 The ${\cal PT}$ symmetric square well
model considered, say, on a finite interval $(-1,1)$ is
characterized by the boundary conditions
 \begin{equation}
 \psi(\pm 1) = 0.
 \label{bc}
 \end{equation}
The general ${\cal PT}$ symmetric piece-wise constant interaction
will be represented here by its most elementary form
 \begin{equation}
 \begin{array}{cc}
 V(x) = i\,Z&  \ \ \ \ {\rm Re}\ x < 0\\
 V(x) = -i\,Z&  \ \ \ \ {\rm Re}\ x > 0.
 \end{array}
 \end{equation}
In the regime with unbroken symmetry the solution of this problem
is virtually trivial\cite{sqw} and obeys the rule
 \begin{equation}
 {\rm Im}\ E_n = 0, \ \ \ \ \ \ n = 0, 1, \ldots\
 \label{rule}
 \end{equation}
provided only that $|Z| < 4.48$. In the other words, all the wave
functions remain ${\cal PT}$ symmetric in this weakly
non-Hermitian regime,
 \begin{equation}
 |\psi_n\rangle = {\cal PT}  |\psi_n\rangle.
 \label{wfrule}
 \end{equation}
In ref.\cite{sqw} the question of what happens beyond
$Z_0^{(crit)} \approx 4.48$ has been skipped as apparently
speculative. One can partially understand the neglect of complex
energies as they mimic the collapse of the system into singularity
in the Hermitian limit\cite{PTHO}. Still, in the light of the new
interpretation and generalization of the ${\cal PT}$ symmetry
\cite{Krakov}, such an interpretation is to be changed. Indeed,
via the concept of pseudo-Hermiticity\cite{Ali}, one can deal with
the wave functions with the broken and unbroken ${\cal PT}$
symmetry on equal footing\cite{pseudo}. Mathematically, a natural
transition point from the real spectrum to the states with complex
energies is provided by the unavoided level crossings, well
illustrated by the symmetric (harmonic\cite{PTHO}) as well as
asymmetric (Morse\cite{Morse}) Laguerre-solvable one-dimensional
oscillators.

\subsection{Construction}

The spontaneous breakdown of ${\cal PT}$ symmetry in square well
beyond the above-mentioned lowest critical coupling $Z_0^{(crit)}$
does not mean that the lowest  pair of the energy levels $E_{0}$
and $E_{1}$ ``merges and disappears" as conjectured (erroneously)
in ref.\cite{sqw}. These two energies rather move in the complex
plain,
 \begin{equation}
 E_0=E-i\,\varepsilon, \ \ \ \ \ \
  E_1=E+i\,\varepsilon, \ \ \ \ \ \ Z > Z_0^{(crit)}
  \label{doublet}
  \end{equation}
and merely the rule (\ref{wfrule}) becomes violated. In this
regime of broken ${\cal PT}$ symmetry, our Schr\"{o}dinger
square-well equation reads
 \begin{equation}
 \psi_n^{''} = \left \{
 \begin{array}{c}
 \left [ k_n^{[+]} \right ]^2\,\psi_n, \ \ \ \ \ x > 0\\
 \left [ k_n^{[-]} \right ]^2\,\psi_n, \ \ \ \ \ x < 0\
 \end{array}
 \right .
 \label{ourSE}
 \end{equation}
and remains easily solvable. Let us pick up just $n=0$ and $n=1$
and abbreviate
 \begin{equation}
 \begin{array}{c}
 \left [ k_0^{[+]} \right ]^2=-E+i\,\varepsilon - i\,Z =\kappa^2=
 (s-i\,t)^2,\\
 \left [ k_1^{[+]} \right ]^2=-E-i\,\varepsilon - i\,Z =\lambda^2=
 (p-i\,q)^2,\\
 \left [ k_0^{[-]} \right ]^2
 =
 -E+i\,\varepsilon + i\,Z =\left[ \lambda^*\right]^2,\\
  \left [ k_1^{[-]} \right ]^2
 =
 -E-i\,\varepsilon + i\,Z =\left[ \kappa^*\right]^2\ .
 \end{array}
 \label{jedna}
 \end{equation}
Then, the general solution of eq. (\ref{ourSE}) may be written in
the well known hyperbolic-function form $\psi \sim a\, \cosh k\,x
+ b\,\sinh k\,x$. Its specification compatible with the boundary
conditions (\ref{bc}) is immediately available,
 \begin{equation}
 \begin{array}{c}
 \psi_0(x)=K_p\,\sinh \kappa\,(1-x), \ \ \ \ \ \ x > 0,\\
 \psi_0(x)=K_n\,\sinh \lambda^*\,(1+x), \ \ \ \ \ \ x < 0,\\
 \psi_1(x)=L_p\,\sinh \lambda\,(1-x), \ \ \ \ \ \ x > 0,\\
 \psi_1(x)=L_n\,\sinh \kappa^*\,(1+x), \ \ \ \ \ \ x < 0.
 \end{array}
 \end{equation}
The necessary continuity of the logarithmic derivatives in the
origin has the form of the four matching conditions at $x=0$,
 \begin{equation}
 \begin{array}{c}
 L_p\,\sinh \lambda = L_n\,\sinh \kappa^*,\\
 \lambda\,L_p\,\cosh \lambda =- \kappa^*\,L_n\,\cosh \kappa^*,\\
 K_p\,\sinh \kappa = K_n\,\sinh \lambda^*,\\
 \kappa\,K_p\,\cosh \kappa = -\lambda^*\,K_n\,\cosh \lambda^*.
 \end{array}
 \end{equation}
Two of them define the coefficients $K_p$ and $L_p$ (in terms of
arbitrary $K_n$ and $L_n$). What remains are the two complex
conjugations of the same complex equation
 \begin{equation}
 \lambda\,{\rm coth} \ \lambda + \kappa^*\,{\rm coth}\ \kappa^*=0.
\label{dve}
\end{equation}
We may summarize: The real and imaginary components
of $\kappa$ and $\lambda$  (i.e., the four unknown parameters
$s$, $t$, $p$ and $q$)  define the real and imaginary part
of the energies in
a way which depends implicitly on $Z
 = pq+st
$,
 \begin{equation}
E = t^2-s^2 = q^2 - p^2,
\ \ \ \ \ \ \varepsilon = pq-st.
 \end{equation}
This enables us to re-parametrize
 \begin{equation}
s=k\,\sinh \alpha, \ \ \
t=k\,\cosh \alpha, \ \ \
p=k\,\sinh \beta, \ \ \
q=k\,\cosh \beta
 \end{equation}
and eliminate
 \begin{equation}
k = \sqrt{\frac{2Z}{\sinh 2\alpha + \sinh 2 \beta}}\ .
 \end{equation}
As a byproduct, one gets the definition
 \begin{equation}
\varepsilon = \frac{k^2}{2}\,( \sinh 2 \beta-\sinh 2\alpha ).
 \end{equation}
Our solutions are completely determined by the two free parameters
$ \alpha$ and $\beta$ from real domain $I\!\!R$. Their values have
to satisfy the two transcendental equations, viz. the real and
imaginary part of eq. (\ref{dve}).

\subsection{Graphical analysis}

The explicit values of the complex conjugate energy doublets have
to be sought by a suitable computer routine. The numerical proof
of their existence and completeness can be delivered easily by
their explicit evaluation using MAPLE\cite{Maple}. A sample of the
underlying real parameters $ \alpha$ and $\beta$ is given in Table
1 for several couplings $Z$ near $Z_0^{(crit)}$.

Similar computation can be performed in any range of $Z$, giving
the second series of roots for $Z > Z_1^{(crit)}$ etc. A sample of
results is presented in Table 2 which lists the roots and the real
part of the energy $E$ near the second critical point of the
symmetry breaking. The value of the second critical coupling
constant is determined as $Z_1^{(crit)} \approx 12.80155$.

The first halves of our Tables document and cross-check the
reliability of the numerical method. They reproduce the first and
second excited state and re-confirm the expected coincidence of
the respective roots $\alpha_{1,2} = \beta_{1,2}$ in the ${\cal
PT}$ symmetric regime. Table 1 improves the estimate of the
critical coupling $Z_0^{(crit)} \approx 4.48$ as obtained in ref.
\cite{sqw}. This estimate is consistent with its alternative
determination in ref.\cite{Quesne} using a direct evaluation of
the pseudo-norm which changes its sign at $Z_0^{(crit)} \approx
4.475$.

\vspace*{1pt}\textlineskip
\section{Summary}
\vspace*{-0.5pt} \noindent
 In the standard, Hermitian quantum
mechanics the observables are represented by operators ${\cal O}$.
Their mean values $\langle \psi | {\cal O}| \psi \rangle$ are
given the well known probabilistic interpretation\cite{Messiah}.
We have noted that at least a part of this scheme can be extended
to certain non-Hermitian and, in particular, ${\cal PT}$ symmetric
operators\cite{Ali}. In the literature, the main source of
interest in this alternative is the possible reality of the
energies of the related bound states. This relationship is not too
robust and the first counterexamples appeared in the very
letter\cite{BB} on the anharmonic potentials $V(x) = m^2x^2 -
(ix)^N$ at the sub-harmonic powers $N < 2$. Moreover, one can
work, alternatively, with the non-standard bra-vectors\cite{what}
 \begin{equation}
 \langle \psi | \to Q \cdot \langle \psi | {\cal P} \equiv \langle
 \langle \psi |, \ \ \ \ \ \  Q = \pm 1
 \label{product}
 \end{equation}
admitting that the norm can be formally
indeterminate\cite{Quesne}. This reflects the non-Hermiticity of
the Hamiltonians and facilitates also the perturbative ${\cal PT}$
symmetric calculations\cite{ix3}.

The latter coinsiderations leave the consequent interpretation of
the ${\cal PT}$ symmetry still
open\cite{JaparidzeK}-\cite{kretschmer}. At the same time, its use
already inspired several studies in field theory\cite{BMa} where
the choice of the symbol ${\cal T}$ indicates an intimate
connection of our symmetry with time reversal.

Within the formalism an increasingly important role is played by
the wave functions which lose the ${\cal PT}$ symmetry
``spontaneously". A byproduct is the complexification
(\ref{doublet}) of the energies. This possibility, whenever
encountered\cite{Mandal,Ahmed}, has been considered ``exotic" in
the recent past. Only after one innovates the bra vectors in
accord with eq. (\ref{product}) it becomes clear that one should
work with the doublets of solutions and that the sign $Q$ in eq.
(\ref{product}) plays the role of a quasi-parity. This concept did
already find a natural extension to more dimensions\cite{heinon}
and to the exactly solvable many-body systems\cite{Tater}.

Let us repeat that our choice of the square-well explicit example
has been dictated by several reasons. Firstly, it enables us to
recover a lot of new analogies between the standard and ${\cal
PT}$ symmetric quantum mechanics. Secondly, in contrast to the
purely numerical studies, the graphical solution of the square
well problem remains entirely transparent. Thirdly, in a way
complementing the illustrative harmonic oscillator example of
ref.\cite{PTHO}, the square well model seems more realistic (or at
least less degenerate) in exhibiting the merger of $E_{2n}$ and
$E_{2n+1}$ at {\em the different} critical couplings $Z_0^{(crit)}
< Z_1^{(crit)} < \ldots$, which form a ``naturally" ordered
increasing sequence.

\section*{Acknowledgements}
 \noindent
M. Z.  appreciates the support by the grant Nr. A 1048004 of GA AS
CR, while G. L. acknowledges the OTKA grant no. T031945.


\newpage


\begin{table}[htbp]
Table 1. {Transition from the ${\cal PT}-$symmetric regime of
ref.\cite{sqw} (with equal roots $\alpha=\beta$) to the
symmetry-breaking solutions with non-equal parameters $\alpha >
\beta$ in the domain of $Z \approx Z_0^{(crit)}$.\\
 }
\begin{tabular}{||c||c|c||}
 \hline \hline
{\rm coupling}\ $Z$
 &$\alpha$& $\beta$\\
 \hline
 \hline
 4&
 0.3879114341&
0.3879114341\\
4.4&
0.3549674685&
0.3549674685
\\                  4.46
                        &0.3395406749
                         &0.3395406749\\
4.47&                          0.3340437385&
                          0.3340437385\\
4.474                          &0.3299988242
                          &0.3299988242\\
4.4748
                         &0.3284804301
                          &0.3284804301
                          \\
                          \hline
4.4754
                          &0.3274947400
                          &0.3244090140\\
4.476
                          &0.3302221779
                          &0.3217353463\\
4.478
                          &0.3344402377
                          &0.3176964766\\
4.48
                          &0.3372106009
                          &0.3151052359\\
4.49
                          &0.3461603724
                          &0.3070500670\\
4.5
                          &0.3523980314
                          &0.3017053291\\
5
                          &0.4640173242
                          &0.2326877241\\
 \hline \hline \end{tabular}
  \end{table}


\begin{table}[htbp]
Table 2. {The emergence of the next series of non-equal roots
$\alpha
> \beta$ in the vicinity of the next critical value of $Z\approx
Z_1^{(crit)}\approx 12. 80155$.
\\
 }

\begin{tabular}{||c||c|c|c||}
 \hline \hline
{\rm coupling}\ $Z$
 &$\alpha$& $\beta$&{\rm energy\ (real\ part)}\\
 \hline
 \hline
12.8 &0.202064800 &0.202064800 & 30.8270139
\\
12.801& 0.201694378 &0.201694378 &
30.8890887
\\
12.8014& 0.201428174 &0.201428174 &
30.9330671
\\
12.8015& 0.201301113 &0.201301113 &
30.9538785
\\
12.80154& 0.201191776 &0.201191776 & 30.9716960
\\
\hline
12.80156& 0.201372370 &0.200912853 &
30.9797156
\\
12.80158& 0.201489113 &0.200796634 &
30.9797173
\\
12.8016& 0.201575523 &0.200710750 &
30.9797190
\\
12.8018 &0.202071953 &0.200219572 &
30.9797353
\\
12.802 &0.202384800 & 0.199911977  &
30.9797517
\\
12.9 &0.220635091 &0.184230076&
                             30.9878260
\\
 13 &0.229622933& 0.177852160&
 30.9961877\\
14 &0.281083389&0.151889620&
                             31.0861845
\\
 \hline \hline \end{tabular}
  \end{table}

\end{document}